\documentclass[conference]{IEEEtran}
\usepackage{booktabs} 
\usepackage{graphicx}
\usepackage{cite}
\usepackage{subfigure}
\usepackage{fancyhdr}
\usepackage{url}

\begin{document}

\title{CAVBench: A Benchmark Suite for Connected and Autonomous Vehicles}

\author{\IEEEauthorblockN{Yifan Wang\IEEEauthorrefmark{1}\IEEEauthorrefmark{2}\IEEEauthorrefmark{4}, Shaoshan Liu\IEEEauthorrefmark{3}, Xiaopei Wu\IEEEauthorrefmark{2}, Weisong Shi\IEEEauthorrefmark{2}}
\IEEEauthorblockA{\IEEEauthorrefmark{1}SKL of Computer Architecture, Institute of Computing Technology, CAS, Beijing, China}
\IEEEauthorblockA{\IEEEauthorrefmark{2}Department of Computer Science, Wayne State University, Michigan, USA}
\IEEEauthorblockA{\IEEEauthorrefmark{3}PerceptIn, California, USA}
\IEEEauthorblockA{\IEEEauthorrefmark{4}University of Chinese Academy of Sciences, Beijing, China}
wangyifan2014@ict.ac.cn, shaoshan.liu@perceptin.io, {\{xiaopei.wu, weisong\}}@wayne.edu}

\maketitle

\thispagestyle{fancy} 
              
\lhead{}   
\chead{}  
\rhead{}  
\lfoot{} 
\cfoot{\thepage}  
\rfoot{} 
\renewcommand{\headrulewidth}{0pt}  
\renewcommand{\footrulewidth}{0pt}

\pagestyle{fancy}  
\cfoot{\thepage} 

\begin{abstract}
Connected and autonomous vehicles (CAVs) have recently attracted a significant amount of attention both from researchers and industry. Numerous studies targeting algorithms, software frameworks, and applications on the CAVs scenario have emerged. Meanwhile, several pioneer efforts have focused on the edge computing system and architecture design for the CAVs scenario and provided various heterogeneous platform prototypes for CAVs. However, a standard and comprehensive application benchmark for CAVs is missing, hindering the study of these emerging computing systems. To address this challenging problem, we present CAVBench, the first benchmark suite for the edge computing system in the CAVs scenario. CAVBench is comprised of six typical applications covering four dominate CAVs scenarios and takes four datasets as standard input. CAVBench provides quantitative evaluation results via application and system perspective output metrics. We perform a series of experiments and acquire three systemic characteristics of the applications in CAVBench. First, the operation intensity of the applications is polarized, which explains why heterogeneous hardware is important for a CAVs computing system. Second, all applications in CAVBench consume high memory bandwidth, so the system should be equipped with high bandwidth memory or leverage good memory bandwidth management to avoid the performance degradation caused by memory bandwidth competition. Third, some applications have worse data/instruction locality based on the cache miss observation, so the computing system targeting these applications should optimize the cache architecture. Last, we use the CAVBench to evaluate a typical edge computing platform and present the quantitative and qualitative analysis of the benchmarking results.
\end{abstract}


\IEEEpeerreviewmaketitle

\section{Introduction}
\label{sec:1intro}

With the rapid development of computer vision, deep learning, mobile communication and sensor technology, the functions of vehicles are no longer limited to driving and transportation, but have gradually become an intelligent, connected, and autonomous system. We refer to these advanced vehicles as connected and autonomous vehicles (CAVs). The evolution of vehicles has given rise to numerous new application scenarios, such as Advanced Driver Assistance Systems (ADAS) or Autonomous Driving (AD)\cite{geiger2012we, sea2016taxonomy}, Internet of Vehicles (IoV)\cite{gerla2014internet} and Intelligent Transportation Systems (ITS)\cite{dimitrakopoulos2010intelligent}, etc. Especially for ADAS/AD scenarios, many industry leaders have recently published their own autonomous driving systems, such as Google Waymo\cite{waymo2018waymo}, Tesla Autopilot\cite{tesla2018tesla}, and Baidu Apollo\cite{baidu2018apollo}. 

Under these scenarios, the CAVs system becomes a typical edge computing system\cite{shi2016edge, shi2016promise}. The CAVs computing system collects sensor data via the CAN bus and feeds the data to on-vehicle applications. In addition, the CAVs system is not isolated in the network, so the CAVs will communicate with cloud servers\cite{liu2017unified}, Roadside Unit (RSU) and other CAVs to perform some computing tasks collaboratively. Much research focusing on the edge computing on CAVs from the application aspect have emerged\cite{kar2017real,lee2017gremlin,qi2017vehicle}. There have also been some pioneer studies about exploring the computing architecture and systems for CAVs. NVIDIA\textsuperscript{\textregistered}DRIVE\texttrademark PX2 is an AI platform for autonomous driving that equips two discrete GPUs\cite{nvidia2018nvidia}. Liu et al. proposed their computing architecture for CAVs, which fully used hybrid heterogeneous hardware (GPUs, FPGAs, and ASICs)\cite{liu2017computer}. Unlike other computing scenarios, edge computing is still a booming computing domain. However, to date, a complete, dedicated benchmark suite to evaluate the edge computing platforms designed for CAVs is missing, both in academic and industrial fields. This makes it difficult for developers to quantify the performance of platforms running different on-vehicle applications, as well as to systematically optimize the computing architecture on CAVs or on-vehicle applications. To address these challenges, we propose CAVBench, the first benchmark suite for edge computing systems on CAVs.

CAVBench is a benchmark suite for evaluating CAVs computing system performance. 
It takes six diverse real-world on-vehicle applications as evaluation workloads, covering four applications scenarios summarized in OpenVDAP: autonomous driving, real-time diagnostics, in-vehicle infotainment and third-party applications\cite{zhang2018openvdap}. The six applications that we have chosen are simultaneous localization and mapping (SLAM), object detection, object tracking, battery diagnostics, speech recognition and edge video analysis. 
We collect four real-world datasets for CAVBench as the standard input to the six applications, which include three types of data: image, audio, and text. CAVBench also has two categories of output metrics. One is application perspective metric, which includes the execution time breakdown for each application, helping developers find the performance bottleneck in the application side. Another is system perspective metric, which we called the Quality of Service - Resource Utilization curve (QoS-RU curve). The QoS-RU curve can be used to calculate the Matching Factor (MF) between the application and the computing platform on CAVs. 
The QoS-RU curve can be considered as a quantitative performance index of the computing platform that helps researchers and developers optimize on-vehicle applications and CAVs computing architecture. 
Furthermore, we analyze the characteristics of the applications in CAVBench. We observe the application information and conclude three basic features of the applications on CAVs. First, the CAVs applications types are diverse, and the real-time applications are dominated in CAVs scenarios. Second, the input data of CAVs applications is mostly unstructured. Third, deep learning applications in CAVs scenarios prefer end-to-end models. Then, we comprehensively characterize the six applications in CAVBench via several experiments. On a typical state-of-the-practice edge computing platform, we have the following conclusions: 

\begin{itemize}
\item The operation intensity of applications in CAVBench is polarized. The deep learning applications have higher floating point operation intensity because the neural networks are their main workloads, which includes plenty of floating point multiplications and additions. As for computer vision applications, the algorithms rely more on mathematical models, which contains lower floating point operation intensity. Hence, the computing platform in CAVs should contain heterogeneous hardware to process these different applications. 

\item Similar to traditional computing scenarios, the applications in CAVBench need high memory bandwidth. That will cause competition for memory bandwidth when multiple real-time applications are running concurrently in real environments. 

\item On average, the CAVBench has a lower cache and TLB miss rate, which means the applications in CAVs scenarios have good data/instruction locality, but for specific workloads, some applications have one or two higher miss rates. Thus, the computing systems targeting these applications should value the optimization of the cache architecture and the data/instruction locality to improve the application performance.

\end{itemize}

Finally, we use the CAVBench to evaluate a typical edge computing platform and present the quantitative and qualitative analysis of the evaluation results of this platform.

The remainder of this paper is organized as follows. In Section \ref{sec:2related}, we discuss the related work. Section \ref{sec:3design} summarizes the methodology for designing CAVBench and presents the overview and detailed components of CAVBench. In Section \ref{sec:4character}, we analyze the characteristics of the applications in CAVBench from different views. The experimental evaluation results of CAVBench are illustrated in Section \ref{sec:5evalu}. Finally, we conclude our work in Section \ref{sec:6conclu}.

\section{Related Work}
\label{sec:2related}
CAVBench is designed for evaluating the performance of the computing architecture and system of connected and autonomous vehicles. In this section, we summarize the related work from two aspects: the CAVs computing architecture and system, and the benchmark suite related to CAVs. 

\subsection{Architecture and System for CAVs}

Junior\cite{montemerlo2008junior} was the first work to introduce a complete system of self-driving vehicles, which included type and location of sensors, as well as software architecture design\cite{levinson2011towards}. Junior presented dedicated and comprehensive information about applications and a software flow diagram for autonomous driving. However, Junior provided less information about the computing system on their self-driving vehicles. 

Liu et al. proposed a computer architecture for autonomous vehicles which fully used hybrid heterogeneous hardware\cite{liu2017computer}. In this work, the applications for autonomous driving were divided into three stages: sensing, perception, and decision-making. They compared the performance of different hardware running basic autonomous driving tasks and concluded some rules to perform different tasks for dedicated heterogeneous hardware. Lin et al. explored the architectural constraints and acceleration of autonomous driving system in \cite{lin2018architectural}. They presented a detailed comparison of accelerating related algorithms using heterogeneous platforms including GPUs, FPGAs, and ASICs. The evaluation metrics included running latency and power, which will help developers build an end-to-end autonomous driving system that meets all design constraints. OpenVDAP\cite{zhang2018openvdap} proposed a vehicle computing unit, which contained a tasks scheduling framework and heterogeneous computing platform. The framework scheduled the tasks to specific acceleration hardware, according to task computing characteristics and hardware utilization. In the industrial field, there are several state-of-the-practice computing platforms designed for CAVs, such as NVIDIA\textsuperscript{\textregistered}DRIVE\texttrademark PX2\cite{nvidia2018nvidia} and Xilinx\textsuperscript{\textregistered}Zynq\textsuperscript{\textregistered}UltraScale+\texttrademark ZCU106\cite{xilinx2018xilinx}.

These projects can be regarded as pioneering research in exploring the computing architecture and systems for connected and autonomous vehicles from different aspects. However, the evaluation method of these systems lacks uniform standards; all the research groups chose application type and implementation from their perspectives. Hence, it is challenging to evaluate and compare these systems fairly and comprehensively.

\subsection{Benchmark Suite Related to CAVs}

There are many classic benchmark suites in the traditional computing field, such as BigDataBench\cite{wang2014bigdatabench} for big data computing, Parsec\cite{bienia2008parsec} for parallel computing and HPCC\cite{luszczek2006hpc} for high-performance computing etc. However, for the computing scenario in CAVs, the benchmark research work is still at the beginning stage and can be divided into two categories according to their contents: datasets and workloads.

KITTI\cite{geiger2012we,geiger2013vision} was the first benchmark datasets related to autonomous driving. It comprised rich stereo image data and 2D/3D object annotated data. According to different data types, it also provided a dedicated method to generate the ground truth and calculate the evaluation metrics. KITTI was built for evaluating the performance of algorithms in the autonomous driving scenario, including but not limited to optical flow estimation, visual odometry, and object detection. There are some customized benchmark datasets for each algorithm, such as TUM RGB-D\cite{sturm2012benchmark} for RGB-D SLAM, PASCAL3D\cite{xiang2014beyond} for 3D object detection and the MOTChallenge benchmark\cite{leal2015motchallenge,milan2016mot16} for multi-target tracking. These kinds of benchmark suites will help us choose the implementations and datesets of CAVBench. 

Another class of related benchmark suites used a set of computer vision kernels and applications to benchmark novel hardware architectures. SD-VBS\cite{venkata2009sd} and MEVBench\cite{clemons2011mevbench} both are system performance benchmark suites based on computer vision workloads in diversified fields. SD-VBS assembled 9 high-level vision applications and decomposed them into 28 common computer vision kernels. It also provided single-threaded C and MATLAB implementations of these kernels. MEVBench focused on a set of workloads related with visual recognition applications including feature extraction, feature classiﬁcation, object detection and tracking, etc. MEVBench provided single- and multi-threaded C++ implementations for some of the vision kernels. However, these two benchmarks are prior works in the field, so they are not targeted toward heterogeneous platforms such as GPUs. SLAMBench\cite{nardi2015introducing} concentrated on using a complete RGB-D SLAM application to evaluate novel heterogeneous hardware. It chose KinectFusion\cite{newcombe2011kinectfusion} as the implementation and provided C++, OpenMP, OpenCL and CUDA versions of key function kernels for different platforms. The RGB-D cameras are more suitable for indoor environments, and the workload type in SLAMBench is single. These efforts are a step in the right direction, but we still need a comprehensive benchmark which contains diverse workloads that cover varied application scenarios of CAVs to evaluate the CAVs system as we mentioned above.

\section{Benchmark Design}
\label{sec:3design}
The objective of developing CAVBench is to help developers determine if a given computing platform is competent for all CAVs scenarios.
This section presents the methodology, overview, and components of our CAVBench.

\subsection{Methodology and Overview}
 
 \begin{figure*}[t]
 	\centering
 	\includegraphics[height=2in]{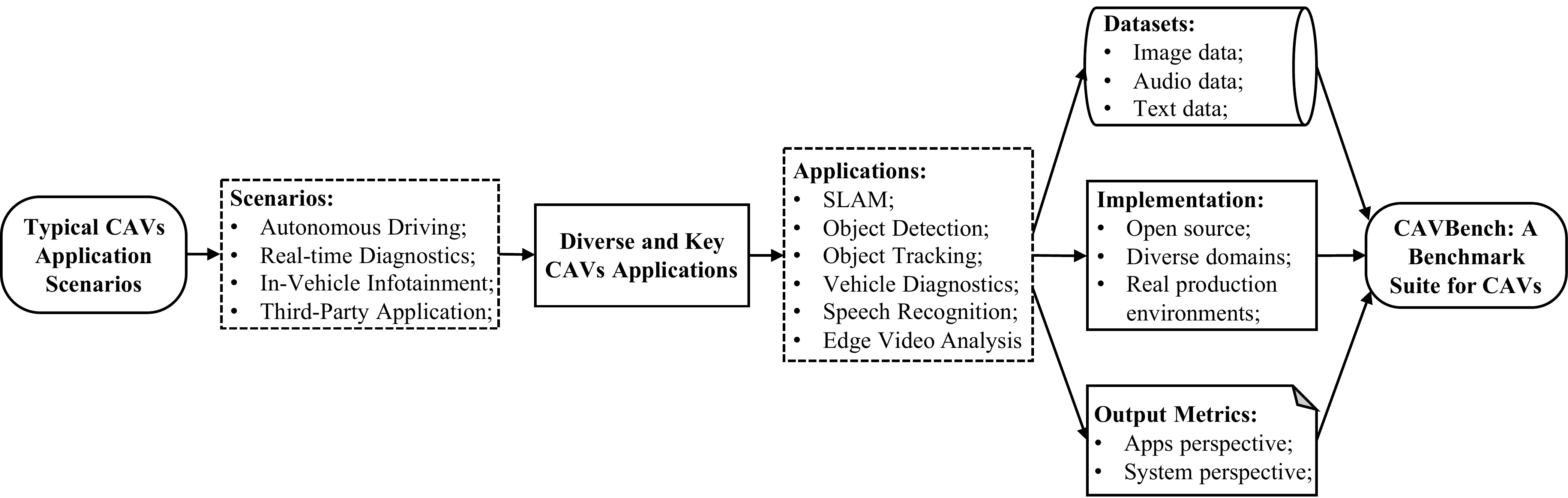}
 	\caption{CAVBench Methodology.}
 	\label{fig:methodology}
 \end{figure*}

Combined with the survey and analysis of the related works on CAVs architectures and systems and existing benchmark suites, we present our methodology for designing CAVBench as shown in Figure \ref{fig:methodology}.
The computing and application scenarios of connected and autonomous vehicles are much different from the traditional domain. 
First, we investigate the typical application scenarios of CAVs. It is well-known that Advanced Driver-Assistant Systems (ADAS) and Autonomous Driving (AD) have already become a dominant application scenario of CAVs\cite{urmson2008autonomous,levinson2011towards,geiger2012we,berger2014autonomous}. In addition to ADAS/AD, OpenVDAP summarizes three other scenarios which are Real-time Diagnostics (RD), In-Vehicle Infotainment (IVI) and Third-Party Application (TApp)\cite{zhang2018openvdap}. Thus, we focus on the exemplary and key applications in each dominant scenario. 

The tasks in the ADAS/AD scenario can be divided into three stages according to their functions: sensing, perception, and decision-making\cite{liu2017computer}. Sensing tasks manage and calibrate the various sensors around the CAVs and provide reliable sensing data to upper-level tasks. Perception tasks take the sensing data as the input and output the surrounding information to the decision-making tasks, which in turn generate a safe and efficient action plan in real time. It can be seen that perception is an important connecting link between sensing and decision-making. The three main perception tasks are simultaneous localization and mapping (SLAM), object detection, and object tracking, which are all visual-based applications. Many studies take them as the vital parts in the autonomous driving pipeline\cite{lin2018architectural,geiger2012we,kato2015open}. Hence, we chose these three applications in the ADAS/AD scenario.

Vehicle system fault diagnostics and prognosis is important for keeping vehicles stable and safe\cite{zhang2009connected}. With the development and widespread use of electric and hybrids vehicles, the health monitoring and diagnostics of Li-ion batteries in these vehicles have received increasingly more attention\cite{zhang2011review}. It is extremely important to monitor and predict the battery status in a real-time fashion, including multiple parameters of each battery cell, e.g., voltage, current, temperature, and so on. Thus, battery diagnostics will be the application chosen in the RD scenario.

The In-Vehicle infotainment (IVI) scenario includes a wide range of applications that provide audio or video entertainment. Compared with manual interaction, the speech-based interaction method reduces the distraction of drivers and ensures driving safety\cite{maciej2009comparison}, so motor companies have increasingly begun to develop their own IVI systems with speech recognition such as Ford\textsuperscript{\textregistered} SYNC\textsuperscript{\textregistered}\cite{ford2018sync}. We chose speech recognition applications for the IVI scenario. 

There are some preliminary projects for the third-party application scenario. PreDriveID\cite{kar2017predriverid} is a driver identification application based on in-vehicle data. It can enhance vehicle safety by detecting whether the driver is registered or not through by analyzing how a driver operates a vehicle. A3\cite{zhang2017enhancing} is an edge video analysis application which uses a vehicle onboard camera to recognize targeted vehicles to enhance the AMBER alert system. It is easier to acquire the data from an onboard camera than the vehicle bus data, and edge video analysis could well be a killer application for edge computing\cite{ananthanarayanan2017real}, so we chose this kind of application for the TApp scenario.

After selecting the six applications, we pay attention to the implementation, datasets and output metrics for the applications. The implementation of each application should be state-of-the-art and representative, ensuring that it can be deployed in a real production environment. We provide real-world datasets for each application which are open source or collected by ourselves to let the applications have a standard input. To give the user a complete understanding of the benchmark results, the output metrics contain two categories: application perspective metric and system perspective metric. These three parts (implementation, datasets and output metrics) form the CAVBench and will be introduced in detail in the next three subsections, respectively.
 
\subsection{Implementation}

The implementation we chose in CAVBench is shown in Table \ref{tab:implementation}. The reasons for choosing these implementations are presented as follows. 

\begin{table*}[h]
\centering
\caption{Overview of Implementation in CAVBench}
\label{tab:implementation}
\begin{tabular}{lllllll}
\hline
Scenario & Application         & App Type   & Implementation                        & Main Workloads        & Data Type         & Data Source \\ \hline \hline
ADAS/AD  & SLAM                & Real-Time          & ORB-SLAM2\cite{kar2017predriverid}    & ORB Extractor and BA  & Unstructured      & Image (Stereo)     \\ 
ADAS/AD  & Object Detection    & Real-Time          & SSD\cite{liu2016ssd}                  & CNNs                  & Unstructured      & Image (Monocular)     \\ 
ADAS/AD  & Object Tracking     & Real-Time          & CIWT\cite{osep2017combined}           & EKF and CRF Model     & Unstructured      & Image (Stereo)     \\ 
RD       & Battery Diagnostics & Offline            & EVBattery                             & LSTM Networks         & Semi-Structured   & Text      \\ 
IVI      & Speech Recognition  & Interactive        & DeepSpeech\cite{hannun2014deep}       & RNNs                  & Unstructured      & Audio     \\ 
TApp     & Edge Video Analysis & Interactive        & OpenALPR\cite{openalpr2018openalpr}   & LBP Feature Detector  & Unstructured      & Image (Monocular)      \\ \hline
\end{tabular}
\end{table*}

\subsubsection{SLAM}
\begin{figure}[h]
 	\centering
 	\includegraphics[width=3.1in]{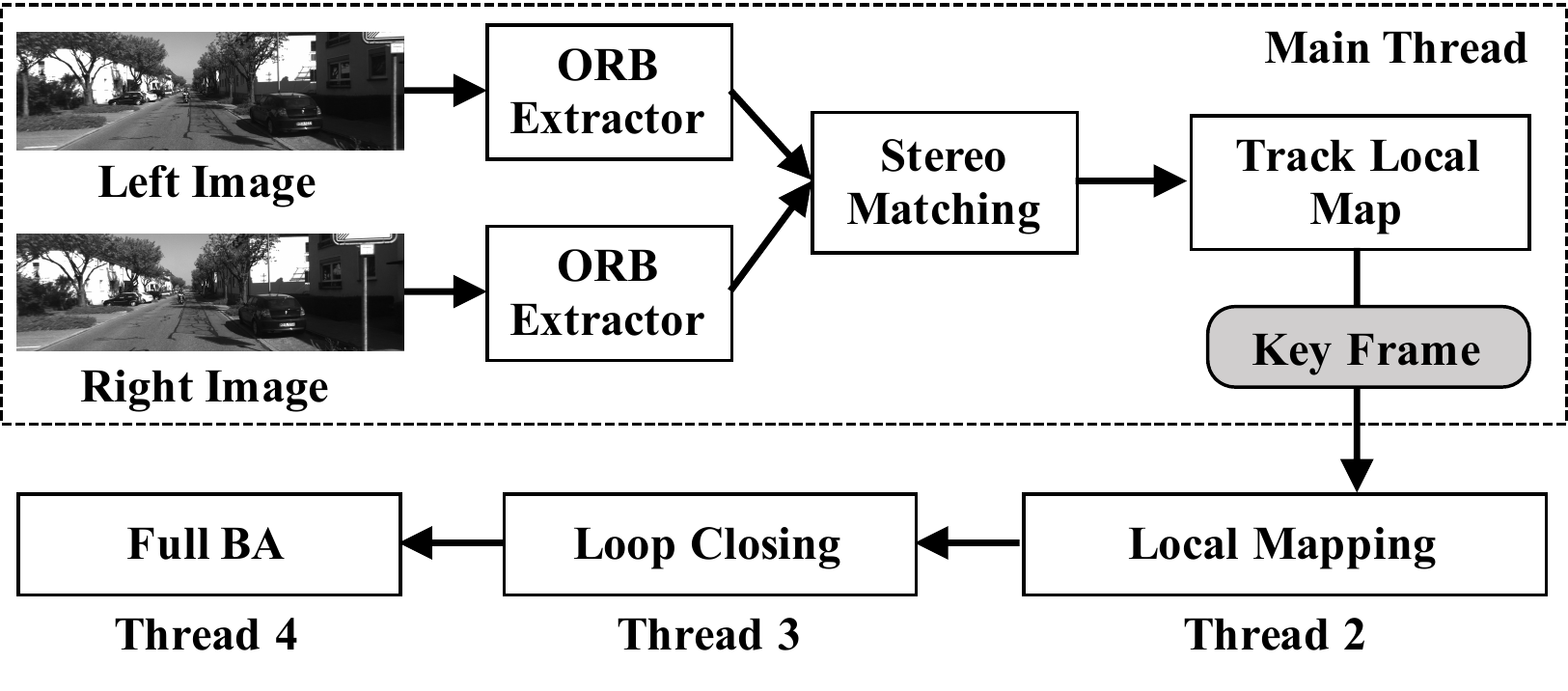} 
 	\caption{Overview of the ORB-SLAM2 Pipeline.}
 	\label{fig:orb_slam2}
 \end{figure}

The simultaneous localization and mapping (SLAM) technique helps CAVs with real-time building a map of an unknown environment and localizing themselves in the map.
 ORB-SLAM2 provides a stereo SLAM method, which has been ranked as the top in KITTI benchmark datasets according to the accuracy and runtime. ORB-SLAM2 is more suitable for large-scale environments than monocular\cite{mur2015orb} and RGB-D SLAM\cite{endres20143}, so we chose it as the SLAM implementation. Figure \ref{fig:orb_slam2} shows the ORB-SLAM2 pipeline. The stereo image stream is fed into the ORB extractor to detect feature points and generate descriptions of the extracted feature points. Then, the main thread attempts to match the current descriptions with the prior map point to localize and generate new keyframes. The local mapping thread manages keyframes to create new map point. The loop closing thread tries to detect and close trajectory loops via the last keyframe processed by the local mapping and creates the fourth thread to optimize a global map by the full Bundle Adjustment (BA).

\subsubsection{Object Detection}
\begin{figure}[h]
 	\centering
 	\includegraphics[width=3.1in]{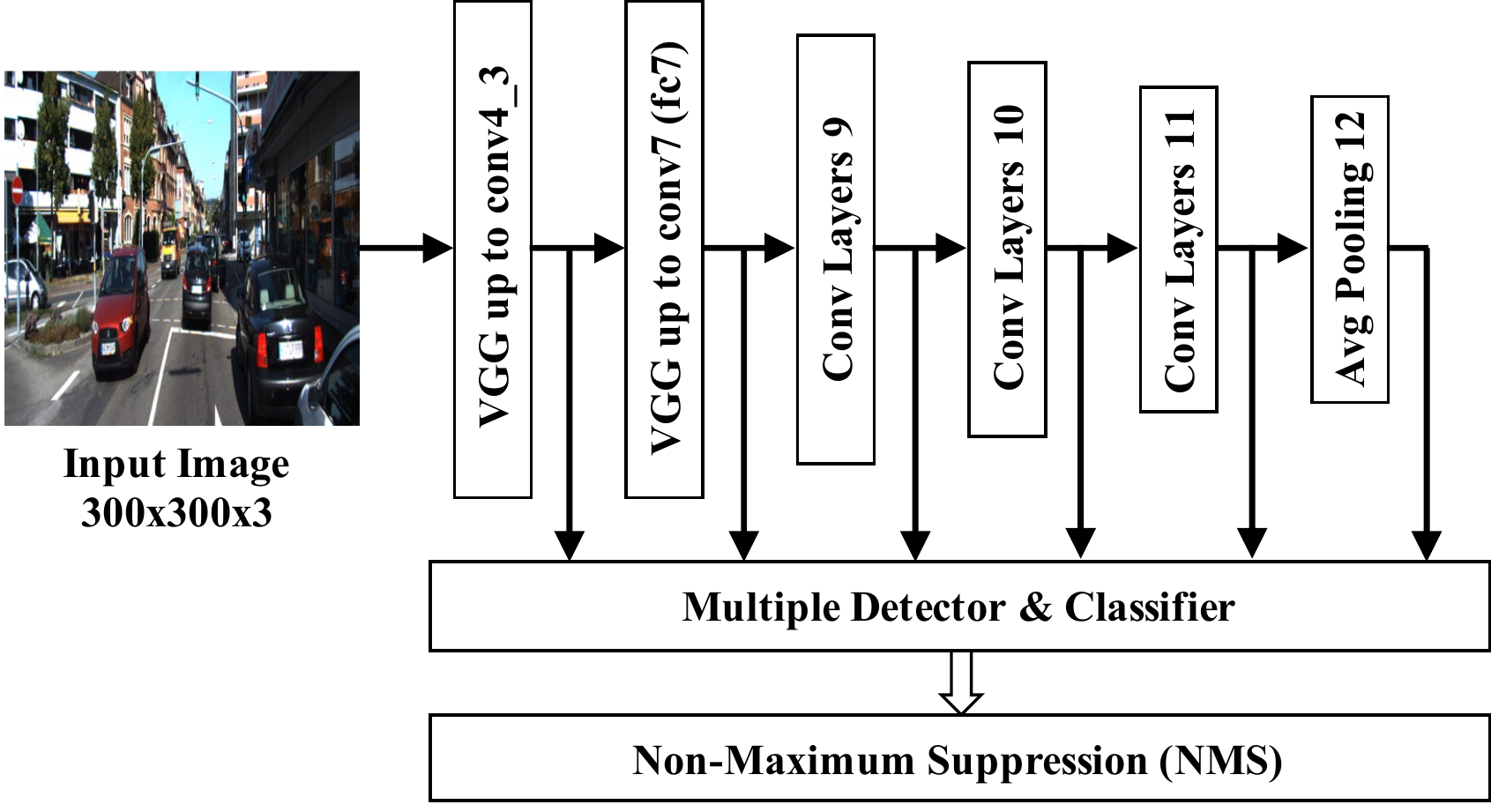} 
 	\caption{Overview of the Single Shot MultiBox Detector Network Structure.}
 	\label{fig:ssd}
 \end{figure}

The visual object category recognition and detection have always been a challenging problem in last decade\cite{everingham2010pascal}. In recent years, the series of algorithms based on Convolutional Neural Networks (CNNs) have become one of the mainstream techniques in the field of object detection\cite{ren2015faster,cai2016unified}, especially in the autonomous driving scenario\cite{chen2016monocular}. Single Shot multibox Detector (SSD)\cite{liu2016ssd} and You Only Look Once (YOLO)\cite{redmon2016you} are kinds of end-to-end CNNs model. Compared with the R-CNN series\cite{girshick2014rich}, they do not hypothesize bounding boxes or resample pixels or features for these hypotheses, which improves the speed for detection and is as accurate as the R-CNN series. The network structure of SDD is shown in Figure \ref{fig:ssd}. SSD uses multiple feature maps from the different stages of the network to perform detection at multiple scales, which is more accurate than the detection by one full connection layer in YOLO. In a word, SSD has higher accuracy and processing speed than other models; hence, we chose SSD as the implementation for object detection.

\subsubsection{Object Tracking}
\begin{figure}[h]
 	\centering
 	\includegraphics[width=3.1in]{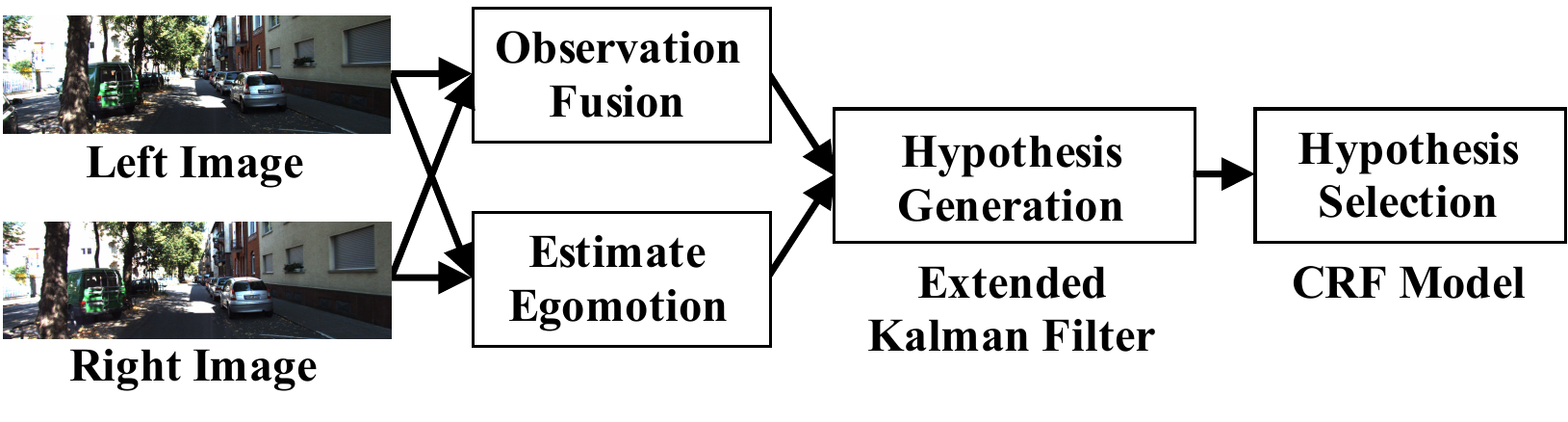} 
 	\caption{Overview of the Combined Image- and World-Space Tracking pipeline.}
 	\label{fig:ciwt}
 \end{figure}

The main goal of object tracking is to ensure that the vehicle does not collide with a moving object, whether a vehicle or a pedestrian crossing the road. We chose Combined Image- and World-Space Tracking (CIWT)\cite{osep2017combined} as the implementation for object tracking, which is an online multiple object tracking application dedicated to the autonomous driving scenario. In KITTI results, CIWT is not the most accurate, but it costs fewer computation resources and less processing time than some algorithms ranked ahead of it. It is more practical in the real-world environment than some offline\cite{geiger20143d} or single target tracking algorithms\cite{held2016learning}. Figure \ref{fig:ciwt} shows the overview of the CIWT pipeline. CIWT uses a stereo image stream to fuse the observation and estimate the egomotion. The observation fusion includes 2D detection and 3D proposals. The tracking process uses these results to generate tacking hypotheses through the Extended Kalman Filter (EKF) and uses the Conditional Random Field (CRF) model to select a high score tacking hypothesis.

\subsubsection{Battery Diagnostics}
\begin{figure}[h]
 	\centering
 	\includegraphics[width=3.1in]{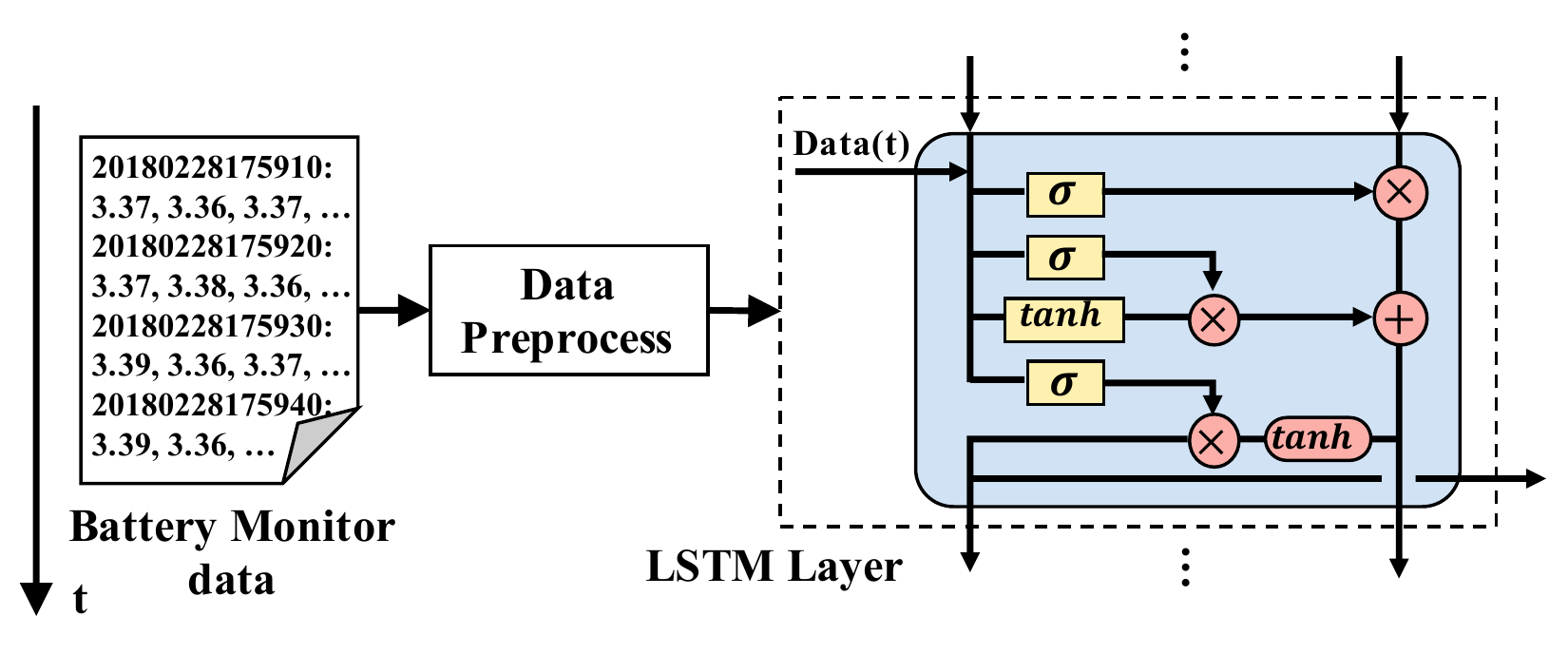} 
 	\caption{Overview of the EVBattery Diagnostics.}
 	\label{fig:evbattery}
 \end{figure}

The devices monitor and log data is a kind of temporal data. Recently, some works leverage Long-Short Term Memory (LSTM) networks to perform failure prediction according to log data for hard drives\cite{dos2017predicting}. LSTM belongs to Recurrent Neural Networks (RNNs), but it has better performance on the long-term prediction task than RNNs. We use the similar method to process EV battery data. We call our implementation as EVBattery Diagnostics and the process is shown in Figure \ref{fig:evbattery}.

\subsubsection{Speech Recognition}
\begin{figure}[h]
 	\centering
 	\includegraphics[width=3.1in]{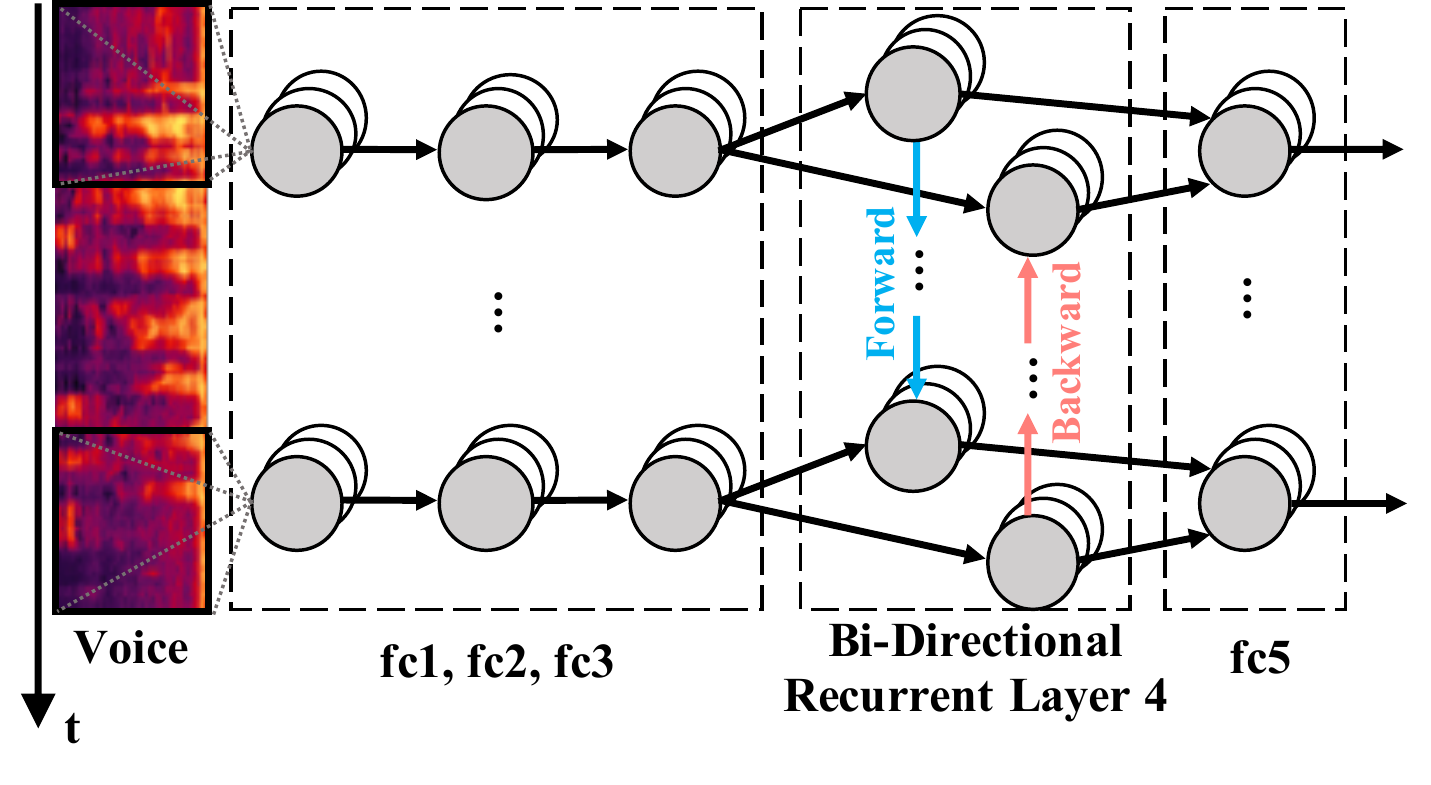} 
 	\caption{Overview of the DeepSpeech Network Structure.}
 	\label{fig:deepspeech}
 \end{figure}

Voice service generally consists of two steps: speech to text and text to intent, with the former being the process of speech recognition. DeepSpeech\cite{hannun2014deep} is an end-to-end speech recognition algorithm based on RNNs. The deep learning method supersedes traditional processing stages in speech recognition systems, such as those that have been hand-engineered. Figure \ref{fig:deepspeech} shows the DeepSpeech network structure. The first three and the fifth layers in DeepSpeech have the basic full connection structure. The fourth layer is a bi-directional recurrent layer which is used to characterize the temporal correlation of the voice data. The evaluation results show that DeepSpeech has less latency and error rates than traditional methods based on Hidden Markov Model (HMM)\cite{maas2014increasing}, especially in noisy environments. Therefore, DeepSpeech is a suitable implementation for speech recognition.

\subsubsection{Edge Video Analysis}
\begin{figure}[h]
 	\centering
 	\includegraphics[width=3.1in]{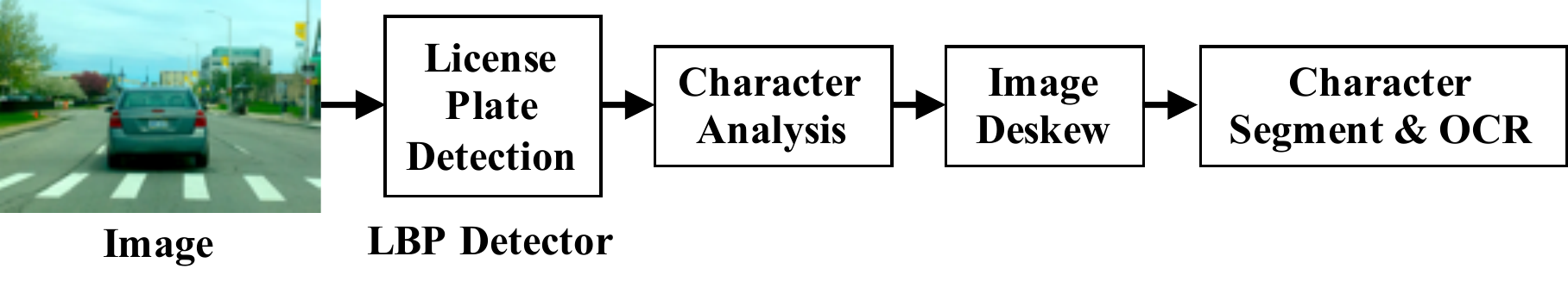} 
 	\caption{Overview of the OpenALPR Pipeline.}
 	\label{fig:openalpr}
 \end{figure}

As we mentioned above, AMBER Alert Assistant (A3) is a typical edge video analysis application that takes OpenALPR\cite{openalpr2018openalpr} as the core workload to detect target vehicles in the video. The OpenALRP pipeline is shown in Figure \ref{fig:openalpr}. OpenALRP is a classic implementation of automatic license plate recognition, including several typical computer vision modules: LBP detector, image deskew and ORC. Therefore, we chose OpenALPR as the implementation of edge video analysis.

\subsection{Datasets}

To guarantee that the evaluation results are similar to running in the real environment, we need to choose a real-world, not a synthetic, dataset for each application. Table \ref{tab:datasets} shows the basic information of the datasets we have chosen. It must be noted that we collected the datasets of Battery Diagnostics and Edge Video Analysis by ourselves because of the lack of relative open source datasets or some datasets did not meet our real-world requirements.

\begin{table}[t]
\centering
\caption{The summary of datasets in CAVBench}
\label{tab:datasets}
\begin{tabular}{|p{2.4cm}|p{2.2cm}|p{2.2cm}|}
\hline
Application         & Datasets                                                  & Data Size and Type                                \\ \hline
SLAM                & KITTI VO/SLAM Datasets\cite{geiger2013vision}             & 21 sequences stereo grayscale image               \\ \hline
Object Detection    & KITTI Object Detection Datasets\cite{geiger2013vision}    & 7518 monocular color image                        \\ \hline
Object Tracking     & KITTI Object Tracking Datasets\cite{geiger2013vision}     & 28 sequences stereo color image                   \\ \hline
Battery Diagnostics & EV Battery Monitor Data                                   & 30 days battery monitor data, 160000 rows \\ \hline
Speech Recognition  & Mozilla Corpus\cite{mozilla2018mozilla}                   & 3995 valid human voice data                       \\ \hline
Edge Video Analysis & Images of Vehicles with License Plates                    & 1000 monocular color image                        \\ \hline
\end{tabular}
\end{table}

\subsubsection{KITTI Datasets\cite{geiger2013vision}}
As mentioned in Section \ref{sec:2related}, KITTI provides rich, open source and real-world image data for different autonomous driving applications. The image characteristics varies, because each dataset focuses on one specific application. Each dataset contains various traffic scenes which can evaluate the application performance comprehensively.

\subsubsection{EV Battery Monitor Data}
There are few datasets that provide vehicle battery monitoring data. We collect the battery monitoring data of an electric vehicle in the real environment for one month. Each record of the data contains 60 items, such as voltage and temperature.

\subsubsection{Mozilla Corpus\cite{mozilla2018mozilla}}
The Mozilla corpus provides 3995 valid common voice files for testing. Each file is a record in which a person reads a common sentence, and records are collected by numerous people reading different sentences. It must be noted that the Mozilla corpus still has some limitations. It was collected in a daily environment, so it may not contain words that are used in the vehicular setting and may not have enough background noise that is likely to be very common in a vehicular environment.

\subsubsection{Images of Vehicles with License Plates}
The images in KITTI datasets could not meet the resolution requirement of performing license plate recognition, and some license plate datasets are not collected by a vehicle onboard camera. Thus, we use the Leopard\textsuperscript{\textregistered} LI-USB30-AR023ZWDRB video camera\cite{leopard2018usb} with 6mm and 25mm lens which was suggested by the Apollo project\cite{baidu2018apollo} to collect image data in real traffic scenes. Each image we provided contains at least one vehicle with its license plate.

\subsection{Output Metrics}
The Output metrics show quantitatively whether the given hardware platform can be used for CAVs scenarios or not. In CAVBench, the output metrics contain two parts: application perspective metric and system perspective metric. 

\subsubsection{Application Perspective Metric}
Like some traditional benchmark suites, the application perspective metric shows the running time of each application. For computer vision applications (ORB-SALM2, CIWT, and OpenALPR), we output the average latency for each module in the applications, and we provide the average and tail latency for deep learning applications (SSD, EVBattery, and DeepSpeech). This metric helps developers optimize the platform in terms of applications.

\subsubsection{System Perspective Metric}
\begin{figure}[t]
 	\centering
 	\includegraphics[width=3.1in]{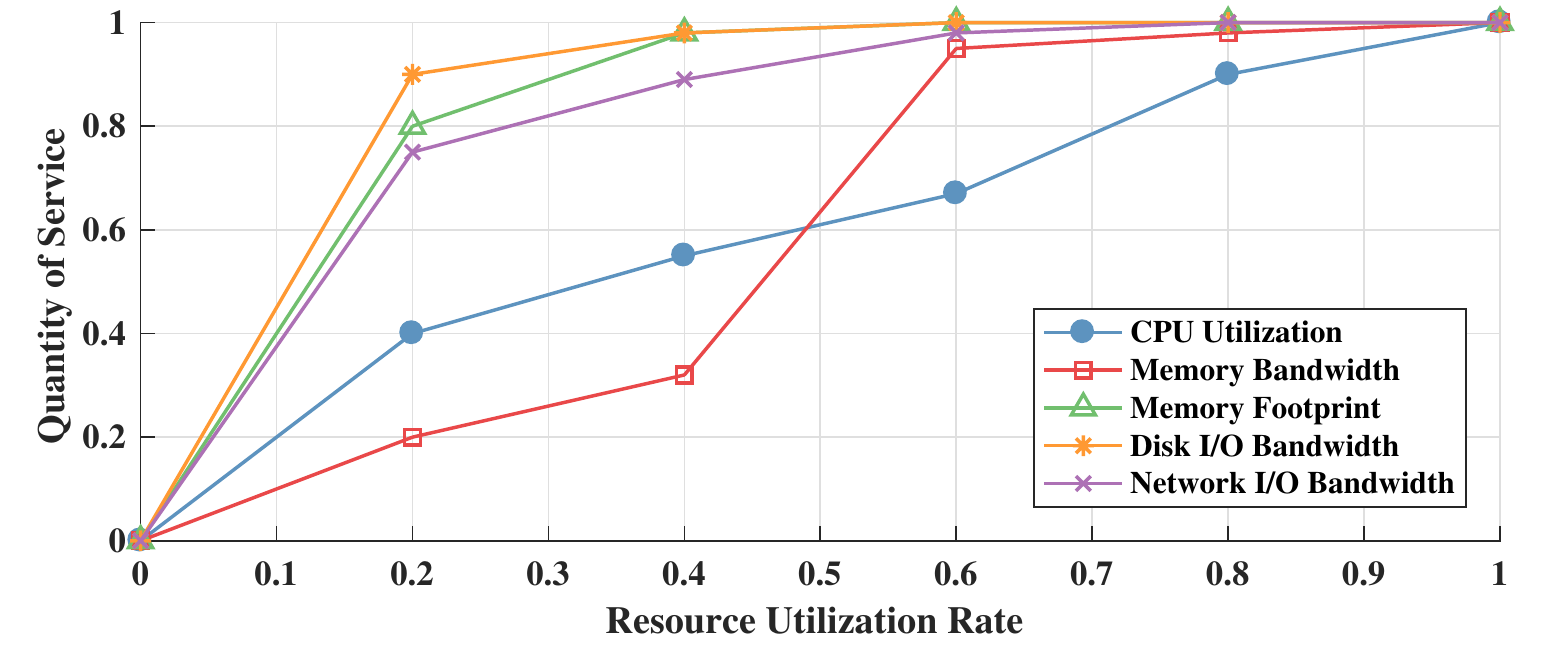} 
 	\caption{An example of QoS-RU Curve.}
 	\label{fig:qos}
 \end{figure}

For the system perspective metric, we call it the quality of service - resource utilization curve (QoS-RU curve). We evaluate the QoS of each application under different system resource allocations and draw the QoS-RU curve for each system resource (CPU utilization, memory footprint, and memory bandwidth, etc.). Figure \ref{fig:qos} shows an example of the QoS-RU curve. We use the area under the curve of each system resource to calculate the Matching Factor (MF) between the application and the platform, indicating whether the platform is suitable for the CAVs application. Following is our approach to calculate the Matching Factor:

We denote the area under the curve of each system resource as $A_i$, and we take the weighted average of each area as the Matching Factor $M$, as Equation \ref{equ:m} shows. 

\begin{equation}
\label{equ:m}
    M=\sum_{i=1}^{n}(w_i\cdot A_i)
\end{equation}
The weight for each resource $w_i$ can be calculated by Equation \ref{equ:wi}, in which the $n$ is the number of system resources we considered.

\begin{equation}
\label{equ:wi}
    w_i=(1-A_i)/\sum_{i=1}^{n}(1-A_i)
\end{equation}
We notice that the $w_i$ is the normalized $1- A_i$. If the $A_i$ is large, the resource $i$ is relatively sufficient for the application. Similarly, if the $A_i$ is small, the resource $i$ has the potential to be the bottle-neck of the platform. Thus, the $A_i$ and its weight have opposite values, which is why we chose the normalized $1- A_i$ as the weight.

A more detailed explanation of the output metrics will be presented in Section \ref{sec:5evalu}.

\section{Benchmark Characterization}
\label{sec:4character}
In this section, we present the detailed description of the experiments of our CAVBench workload characterization analysis.

\subsection{Overview}
Before we did the characterization analysis experiments, we first observed the information of the six applications presented in Table \ref{tab:implementation} to conclude some basic features of the applications in CAVs scenarios. The observations are described as follows:

\subsubsection{Real-Time Applications are Dominant}
The application types in the CAVs computing scenarios are diverse, including real-time, offline and interactive. This diversity corresponds to cloud computing and the big data computing domain\cite{wang2014bigdatabench}. In contrast to traditional computing fields, real-time applications are dominant in CAVs, and they all belong to the ADAS/AD scenario. Furthermore, the applications in other scenarios are offline and interactive. That explains why ADAS/AD applications always have the highest priority. 

\subsubsection{Unstructured Data Type}
The input data type of CAVs applications is mostly unstructured data. As we mentioned above, CAVs is a typical embedded and edge computing system, and it deploys various sensors to collect information from the real physical world, and uses the information (data) to execute computation tasks. Therefore, the input data is generally unstructured, such as images and audios. Even for vehicle monitor data, it usually has little structural constraints, so we consider it as semi-structured data. As for cloud computing, there are still some classic workloads that use the structured data, for example, the relational query operation.

\subsubsection{End-to-End Deep Learning Workloads}
According to the classification of the main workloads in CAVBench, we find that the workloads in CAVs all belong to computer vision, machine learning, and the deep learning domain because the main functions of the applications in CAVs are detection, recognition, and prediction. In addition, due to the limitation of latency, deep learning workloads in CAVs choose end-to-end models, which means that
with the exception of deep neural networks, there are no other processes between input and output; this improves the running speed while not decreasing the algorithm accuracy.

\textbf{[Insights]} 
These three observations can be regarded as the basic characteristics of the applications in the CAVs computing scenario. We can conclude some insights regarding the CAVs computing system and applications design. First, real-time applications have the highest priority in real production environments, so the CAVs system should contain a task scheduling framework to ensure that the real-time applications can be allocated with enough computing resources. Second, preprocessing the unstructured data consumes more time, so a hardware accelerator aimed at transforming the unstructured data to structured will be a benefit to the performance of the whole CAVs system. Third, the end-to-end deep learning algorithm reduces the number and frequency of data movements (main memory to GPUs memory), decreasing the processing latency. Thus, this kind of algorithm will be more suitable for CAVs applications.

\subsection{Experiments Configurations}

\begin{table}[t]
\small
\centering
\caption{Edge Computing Platform Configurations}
\label{tab:edge}
\begin{tabular}{|c|c|}
\hline
Platform           & Intel FRD                     \\ \hline
CPU                & Intel Xeon E3-1275 v5                \\ \hline
Number of Sockets  & 1                                                 \\ \hline
Core(s) per Socket & 4                                               \\ \hline
Thread(s) per Core & 2                                                  \\ \hline
Architecture       & X86\_64                                      \\ \hline
L1d cache          & 4$\times$32KB                                            \\ \hline
L1i cache          & 4$\times$32KB                                         \\ \hline
L2 cache           & 4$\times$256KB                                         \\ \hline
L3 cache           & 8MB                                             \\ \hline
Memory             & 32GB SODIMM 3122 MHz                   \\ \hline
Disk                & 256GB PCIe SSD                   \\ \hline
\end{tabular}
\end{table}

To obtain insights regarding application characteristics in CAVBench, we ran the six applications in a typical edge device, and use the Linux profiling tool Perf to collect the behaviors of the applications at the architecture level. We chose Intel\textsuperscript{\textregistered} Fog Reference Design (FRD) as the experiment platform, which has one Xeon\textsuperscript{\textregistered} E3-1275 v5 processor equipped with 32GB DDR4 memory and 256GB PCIe SSD. The processor has four physical cores, and hyper-threading is enabled. Other detailed information of the platform is shown in Table \ref{tab:edge}. The operating system is Ubuntu 16.04 with Linux kernel 4.13.0. The deep learning applications are built on TensorFlow 1.5.0, and some visual modules are implemented on OpenCV 3.3.1. To acquire the pure and original characteristics of the applications, the platform is not equipped with heterogeneous devices. Each application executes for 500 seconds, sequentially processing different data. 

\subsection{Operation Intensity}

\begin{figure}[b]
 	\centering
 	\includegraphics[width=3.1in]{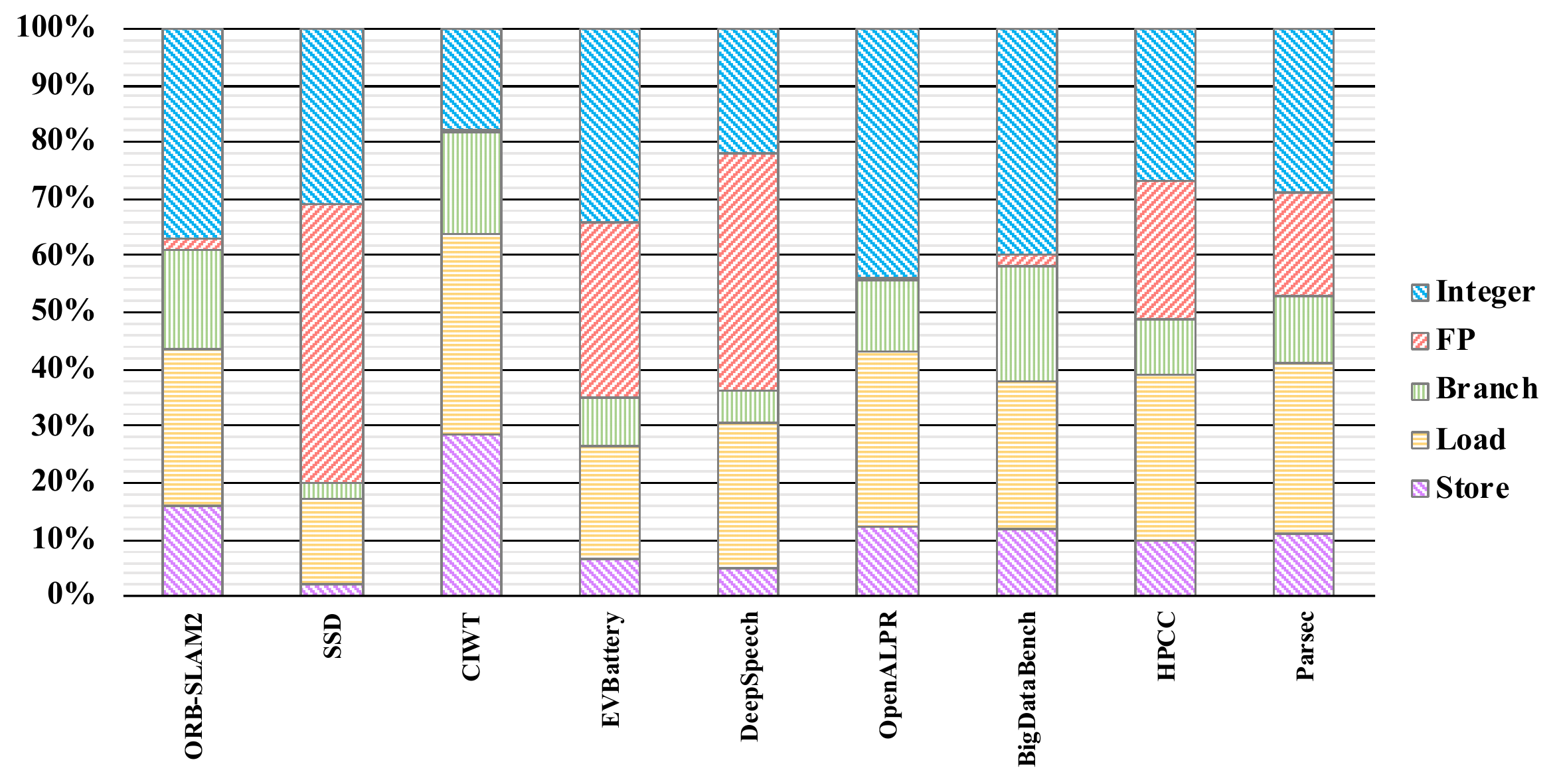}
 	\caption{Instruction Breakdown.}
 	\label{fig:inst}
\end{figure}
First, we analyzed the operation intensity of CAVBench via the instruction breakdown of each application in CAVBench. As shown in Figure \ref{fig:inst}, the distribution of the instructions is diverse, even polarized. The difference is mainly due to floating point instruction (FL). The average proportion of floating point instruction in CAVBench is $20.77\%$, and the average ratio of integer instructions (Int) to FL is $24.44$. However, for each application, the minimum and maximum FL proportion is $0.38\%$ (CIWT) and $48.89\%$ (SSD), and the minimum and maximum Int/FL ratio is $0.63$ (SSD) and $79.59$ (OpenALPR). In addition, the average FL proportion for BigDataBench, HPCC and Parsec are $2.12\%$, $24.11\%$, and $18.25\%$, respectively. The instruction distribution in CAVBench is similar to HPCC, Parsec according to the average values, but the distribution is also polarized when investigating specific applications in CAVBench. In contrast, the instruction distribution of each workload in the traditional benchmark is similar, such as the results presented in BigDataBench\cite{wang2014bigdatabench}. 

\begin{figure}[t]
 	\centering
 	\includegraphics[width=3.1in]{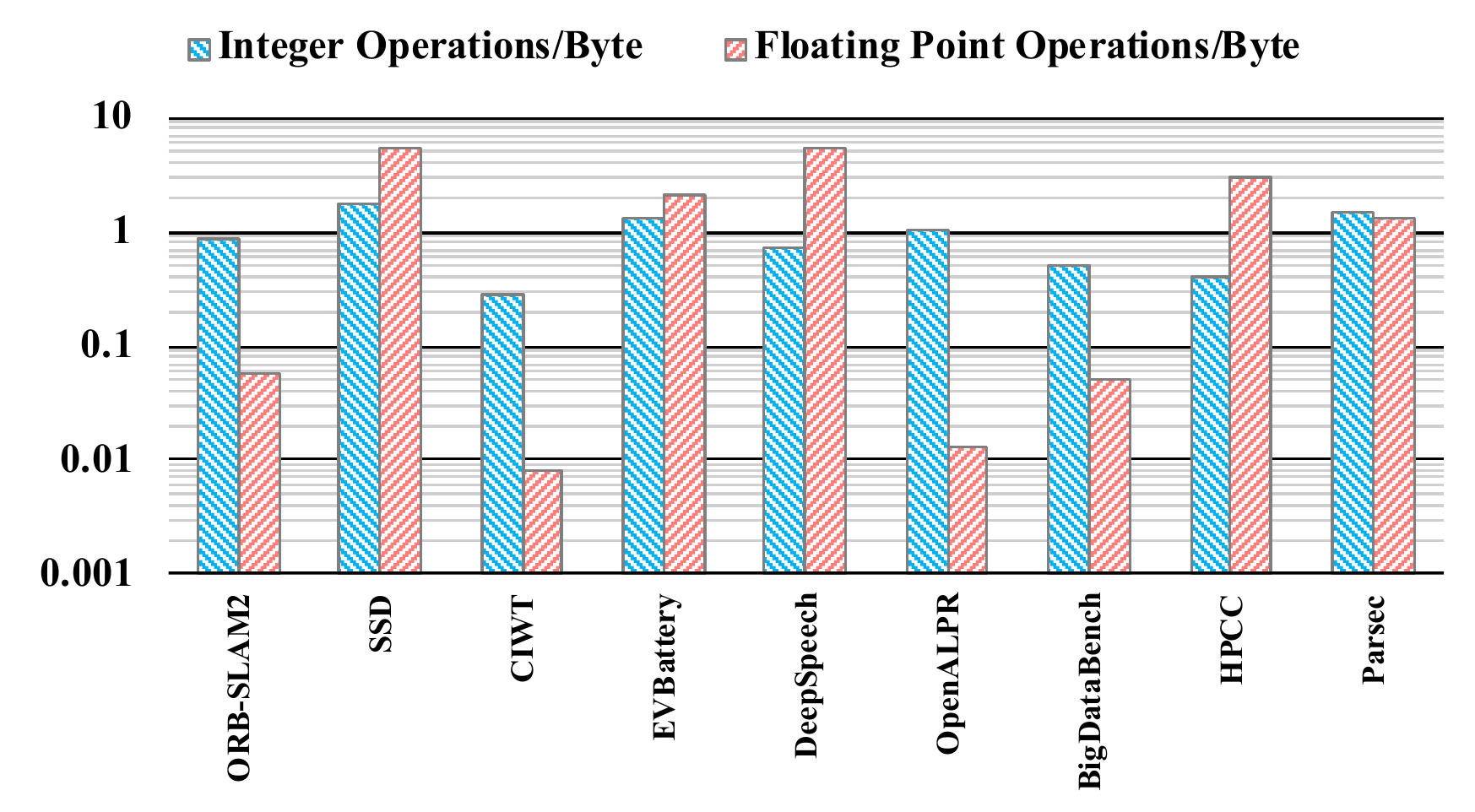} 
 	\caption{Integer and Floating Point Operation Intensity.}
 	\label{fig:operation}
\end{figure}

In order to characterize the computation behaviors, we calculated the ratio of computation to memory access for each application, which represents the integer and floating point operation intensity. As shown in Figure \ref{fig:operation}, the floating point operation (FLO) intensity of each application in CAVBench is still polarized, and the minimum and maximum values are $0.0079$ (CIWT) and $5.46$ (SSD),respectively, which differ by about three orders of magnitude. As for integer operation (IntO) intensity, the minimum and maximum values are $0.28$ (CIWT) and $1.80$ (SSD), and the IntO intensity for BigDataBench, HPCC and Parsec are $0.52$, $0.43$ and $1.50$, respectively. Hence, the IntO intensity of CAVBench is almost in the same order of magnitude as those of the other benchmarks. 

\textbf{[Insights]} 
According to operation intensity experiments, we can draw an important conclusion: the operation intensity of applications in the CAVs scenario is polarized. Deep learning applications, such as SSD and DeepSpeech, have higher floating point operation intensity, which is similar to the workloads in the high-performance computing domain. That is because the neural networks are the main workloads in deep learning applications, which includes a large number of matrix operations, causing plenty of floating point multiplications and additions. As for computer vision applications, the algorithms rely more on mathematical modeling, and the computation is not so high; thus, they have lower floating point operation intensity. This phenomenon explains why modern CAVs computing platforms leverage heterogeneous hardware to accelerate some tasks. The state-of-the-practice CPUs provide SSE, AVX instructions for floating point operations, but the performance does not yet match that of the GPUs.

\subsection{Memory Behavior}
Memory is a very important part of the computer system; the memory wall problem exists in many computing domains. Therefore, we further investigated the memory behaviors of CAVBench. Our experiments included three parts: memory bandwidth, memory footprint, and cache behavior, which we discuss below:

\begin{figure}[b]
    \centering
 	\includegraphics[width=3.1in]{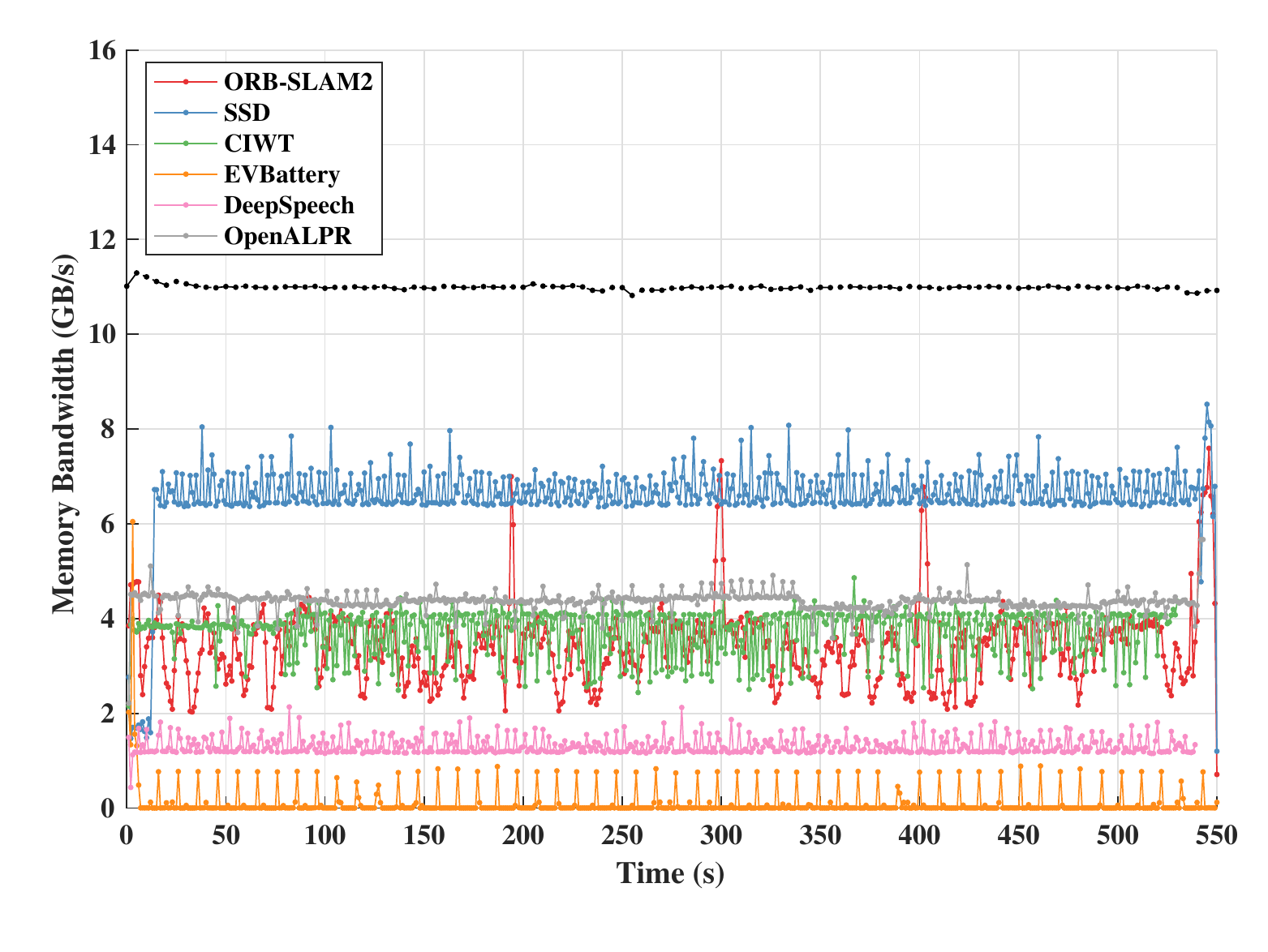} 
 	\caption{Memory Bandwidth Behaviors.}
 	\label{fig:mem_bandwidth}
\end{figure}
Due to the restriction of the hardware function in the processor, we cannot monitor memory bandwidth directly, so we use Perf tools to acquire the indirect memory bandwidth. The measuring method will lead to some errors, but it can still help us analyze the memory access behaviors of each application in CAVBench. 

The real-time memory bandwidth of each application is shown in Figure \ref{fig:mem_bandwidth}. The black line is the memory bandwidth upper limit of our experiment platform, whose average value is $10.98 GB/s$. The applications in CAVBench sequentially process the data in the same size and type, so they all have stable memory bandwidth, as shown in Figure \ref{fig:mem_bandwidth}. The minimum bandwidth is $1.01 GB/s$ (EVBattery), which is $9.20\%$ of the bandwidth upper limit, and the maximum value is $6.57 GB/s$ (SSD), which is $59.84\%$ of the upper limit. According to this observation, we can see that the applications in the CAVs scenario require the high memory bandwidth, and the bandwidth may become the bottleneck of the performance in the real environment. When several CAVs applications run concurrently, they will compete for memory bandwidth resources, leading to higher latency for each application. We observe that the memory bandwidth of ORB-SLAM2 has four peak values during the running time, which reaches $7.21 GB/s$ on average. This is due to the loop closing module performing full BA when a loop is detected in the trajectory. This kind of memory access burst will cause interference in the performance of other applications, especially the tail latency.

\begin{figure}[b]
    \centering
 	\includegraphics[width=3.1in]{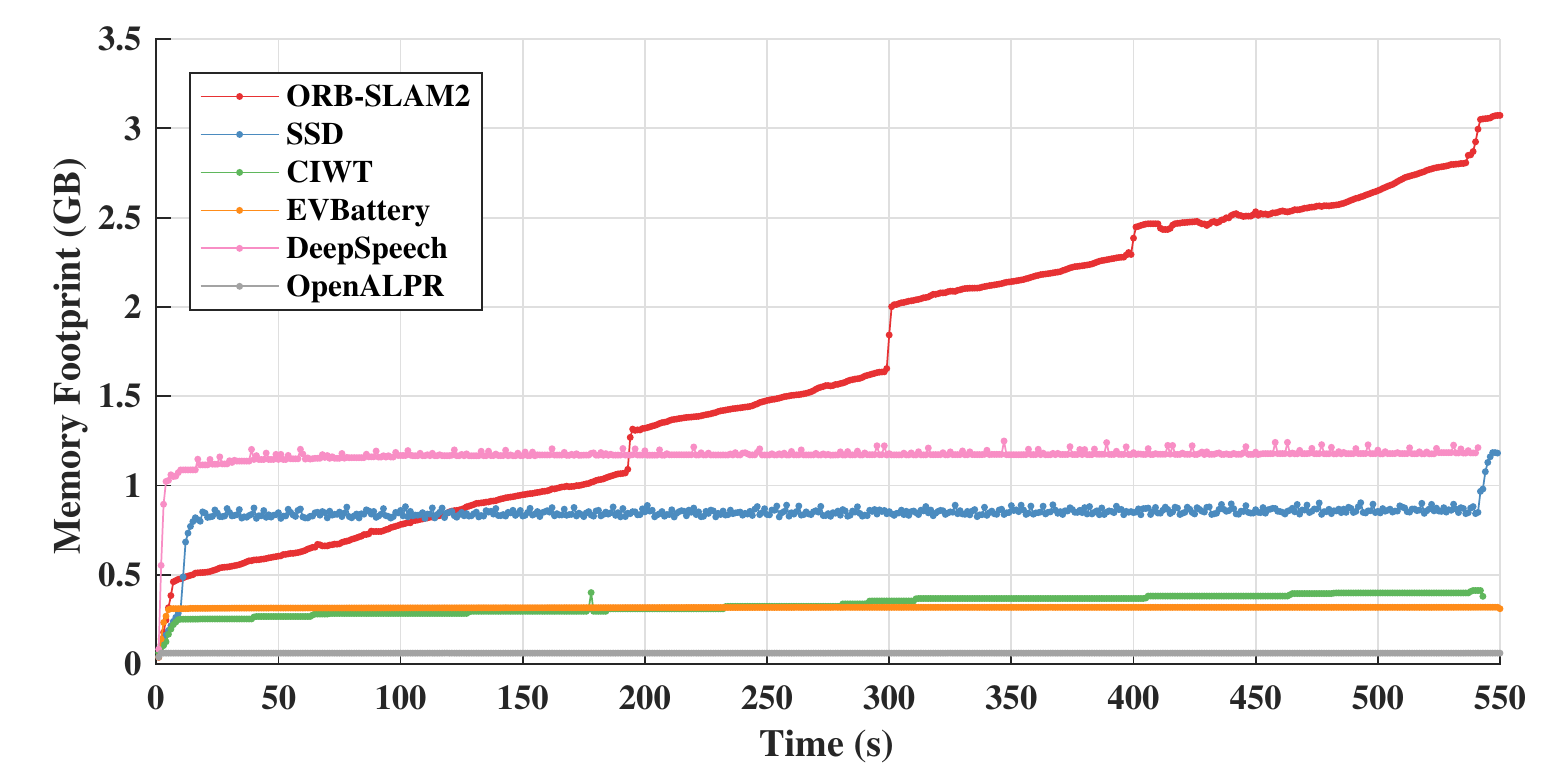} 
 	\caption{Memory Footprint Behaviors.}
 	\label{fig:mem_footprint}
\end{figure}

Furthermore, we investigated the resident memory (memory footprint) behavior of CAVBench. The experiment results are shown in Figure \ref{fig:mem_footprint}. Except for ORB-SLAM2, other applications have stable memory footprint. As we mentioned above, the total memory of the experiment platform is $32GB$. With the exception of ORB-SLAM2, the lowest memory footprint is $0.061GB$ (OpenALPR), which is $0.19\%$ of total memory, and the highest memory footprint is $1.17GB$ (DeepSpeech), which is $3.66\%$ of total memory. The reason for the continued increment of ORB-SLAM2 memory footprint is the application continues to generate new map points and the point data stores in the memory. The four jumps of the memory footprint are also caused by the full BA, which corresponds to the observation in memory bandwidth experiment. The maximum memory footprint of ORB-SLAM2 is $3.07GB$ ($9.59\%$). We can conclude that the applications in the CAVs scenario consume less memory footprint, and the large capacity memory is available for the edge computing platform, so the CAVs application performance will not be constrained by the memory footprint.

\begin{figure}[t]
    \centering
    \subfigure[\label{fig:cache1}Cache Behaviors]{\includegraphics[width=3.1in]{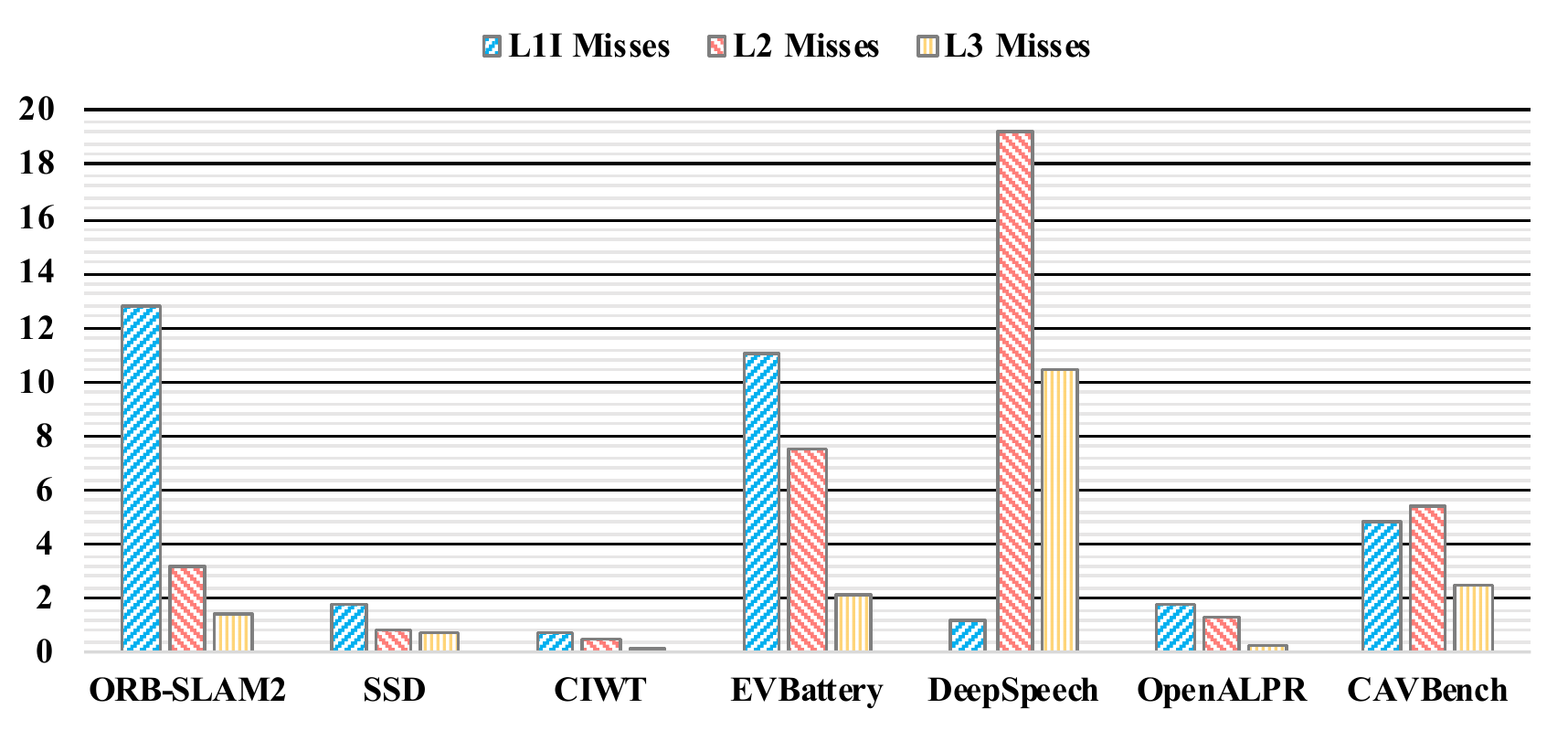}}
    \subfigure[\label{fig:cache2}TLB Behaviors]{\includegraphics[width=3.1in]{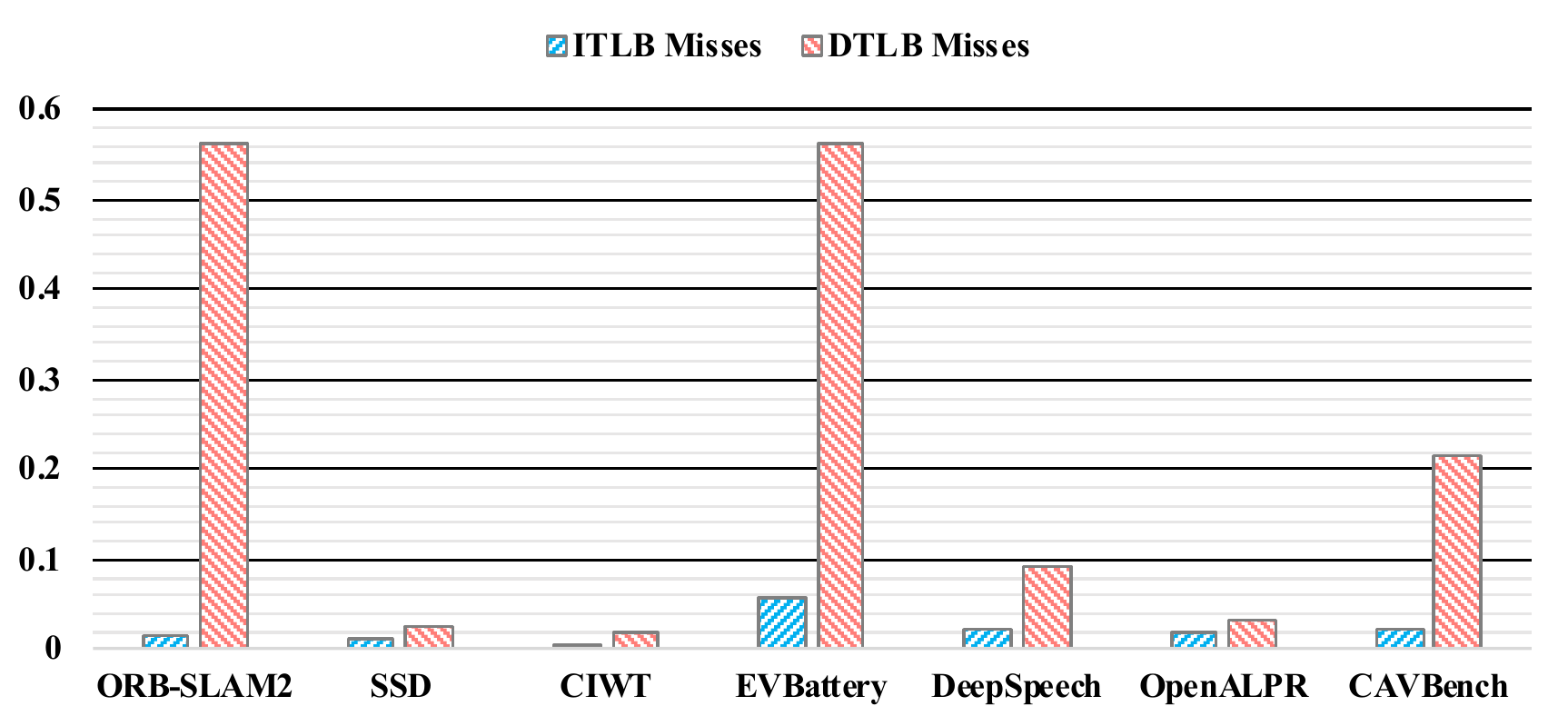}}
    \caption{Cache and TLB Behaviors.}
    \label{fig:cache}
\end{figure}

Finally, we investigated the cache behaviors of CAVBench to see that whether the memory hierarchy architecture in the state-of-the-practice edge platform is proper for CAVs applications. The cache behavior and TLB behavior of each application are shown in Figure \ref{fig:cache}. The CAVBench average L1i, L2 and L3 cache MPKI (Misses Per Kilo Instructions) are $4.86$, $5.44$, and $2.51$, respectively, which are almost in the same order of magnitude as HPCC ($0.41$, $5.59$, and $4.22$) and Parsec ($3.51$, $7.25$, and $3.37$), respectively. The TLB behavior of CAVBench is also similar to HPCC and Parsec. According to this observation, we find that the applications in CAVBench have good data and instruction locality. These characteristics differ from the big data computing, which has a huge code size and deep software stack leading to higher MPKI.

Focusing on specific applications, ORB-SLAM2, EVBattery, and DeepSpeech have a higher MPKI than the other three. The ORB-SLAMS have high L1i MPKI and DTLB MPKI. We infer that this phenomenon is still caused by the periodic loop detection operation and irregular full BA operation. The loop closing module queries the local map database periodically to detect the potential trajectory loop. The DLTB has less capacity to store all the page tables of the local map database, causing the high DTLB miss rate, and the irregular full BA operation interferes with the instruction locality, increasing the L1i cache miss rate. As for EVBattery and DeepSpeech, we infer that the RNNs structure leads to a high cache miss rate. The convolution operations in CNNs make the data and instruction localized, which is why SSD does not have a high cache miss rate, but the RNNs do not have such convolution operations.

\textbf{[Insights]} 
According to the memory experiments, we can draw some important conclusions: First, applications in the CAVs scenario consume high memory bandwidth, which will be a performance bottleneck when multiple applications run concurrently in real environments. Second, the memory footprint of each application takes a very low proportion of the total memory in the state-of-the-practice edge computing platform. Third, on average, the applications in the CAVs scenario have good data and instruction locality. This characteristic is similar to the workload in high-performance computing and the parallel computing domain. As for specific applications, the SLAM and RNNs model based applications have a higher probability to increase the cache and TLB miss rate. Therefore, with the CAVs computing system paying more attention to these applications, we should focus on the optimization of the cache architecture and the application data/instruction locality.

\section{Evaluation}
\label{sec:5evalu}

We use the CAVBench to evaluate our typical edge computing platform; its configurations are presented in Section \ref{sec:4character}. The evaluation results are presented in this section.

\subsection{Latency Results}

\begin{figure*}[t]
    \centering
    \subfigure[\label{fig:eval_orbslam2}ORB-SLAM2]{\includegraphics[height=1.35in]{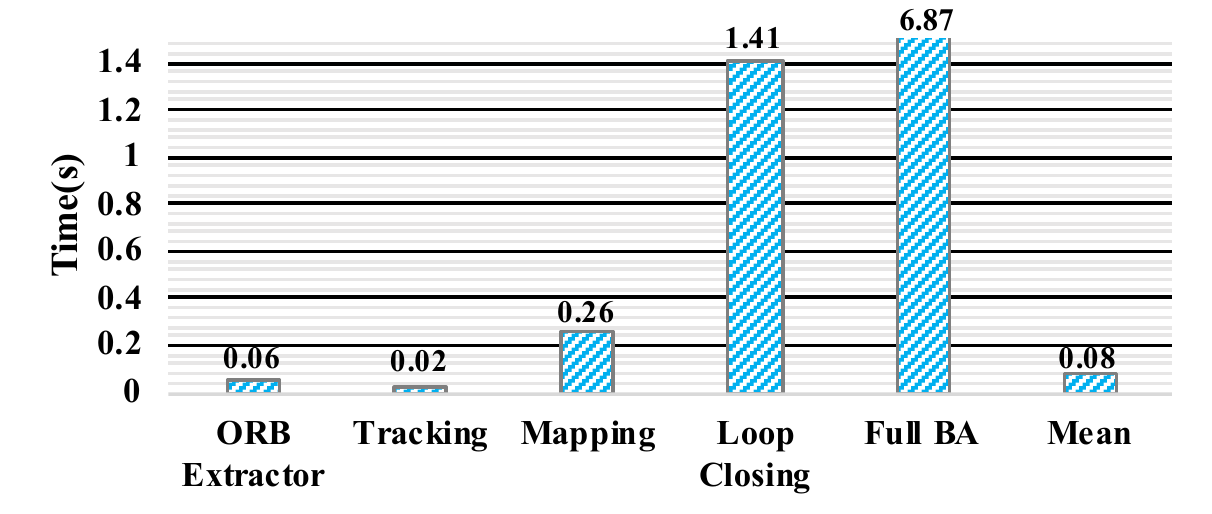}}
    \subfigure[\label{fig:eval_ciwt}CIWT]{\includegraphics[height=1.35in]{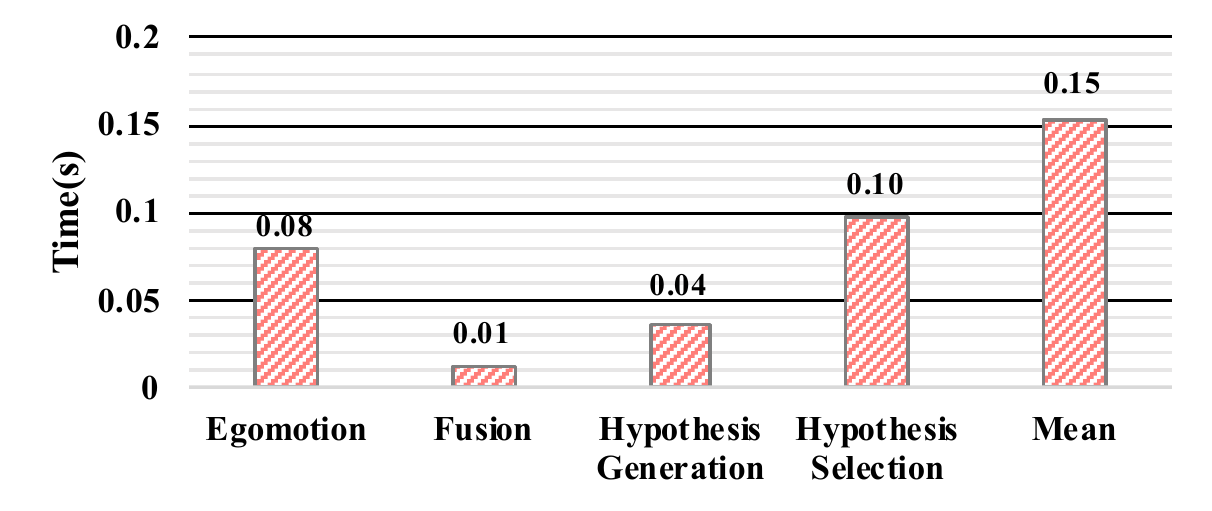}}
    \subfigure[\label{fig:eval_openalpr}OpenALPR]{\includegraphics[height=1.35in]{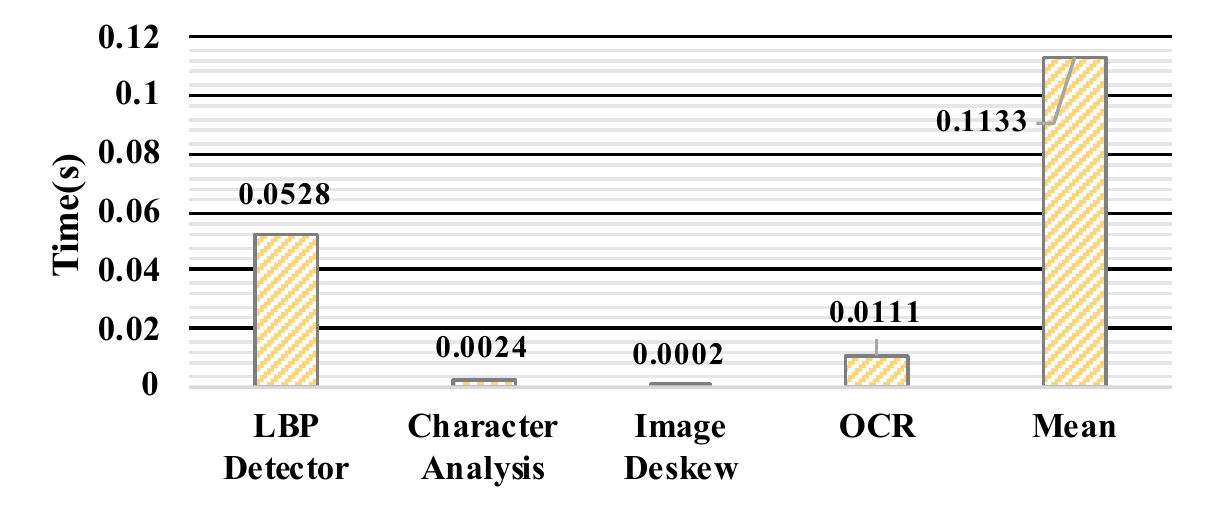}}
    \subfigure[\label{fig:eval_deep}Deep Learning Applications]{\includegraphics[height=1.35in]{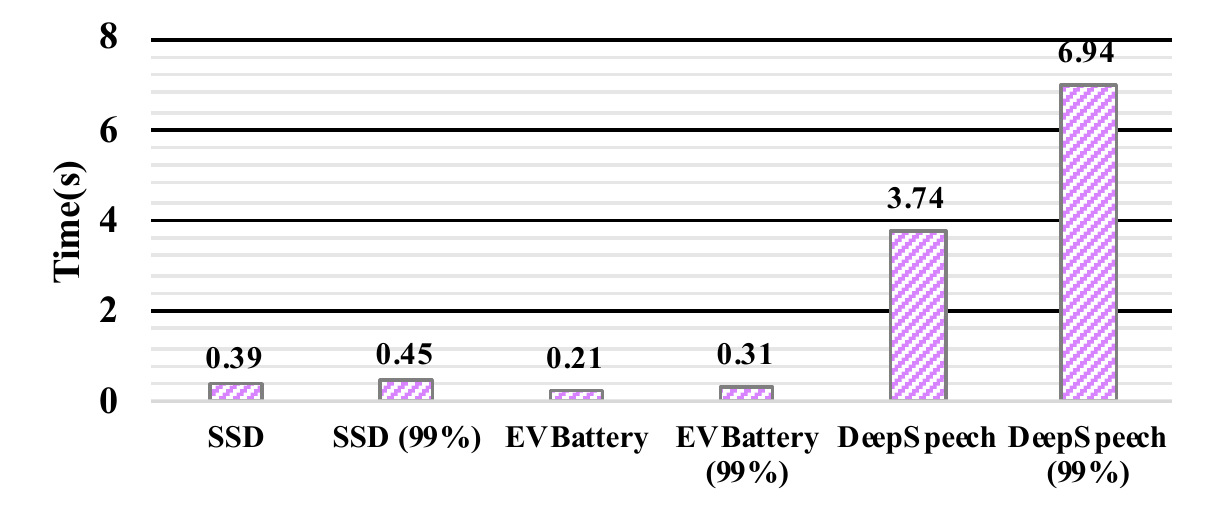}}

    \caption{Application Latency Results on Intel FRD.}
    \label{fig:latency}
\end{figure*}

First, we present the application perspective metric (latency) of the platform in Figure \ref{fig:latency}. According to the results, we find this platform has a good performance in terms of the computer vision applications; the average FPS (Frame Per Second) of ORB-SLAM2, CIWT and OpenALPR are $12.5$, $6.67$ and $8.83$, respectively. The processing speed of these applications is near-acceptable in the real environment. Please note that the latency of Mapping, Loop Closing, and Full BA is quite high, but these modules are not executed for each frame, and they run in different threads, so the latency of these functions is negligible to the average latency.

However, the performance of deep learning applications is not as good as computer vision applications. The FPS of SSD is only $2.55$, which is unacceptable in the real autonomous driving scenario, and the average latency of EVBattery and DeepSpeech are $0.21s$ and $3.74s$, respectively. The latency of DeepSpeech is much longer than user-perceived QoS in an interactive system. Because the input data source of EVBattery is text, the data batch size is only one-hour monitor data, and the neural networks scale is smaller (one layer LSTM), so it has less running latency, which is enough for a real environment. As for SSD and DeepSpeech, they both have a deep and large-scale network structure, and the input data source is image and audio (Unstructured data). The platform is incompetent for these kinds of applications since it is not equipped with heterogeneous hardware to acceleration. 

\subsection{QoS-RU Curve Case Study}

\begin{figure}[h]
    \centering
    \subfigure[\label{fig:curve_orbslam2}ORB-SLAM2]{\includegraphics[height=1.3in]{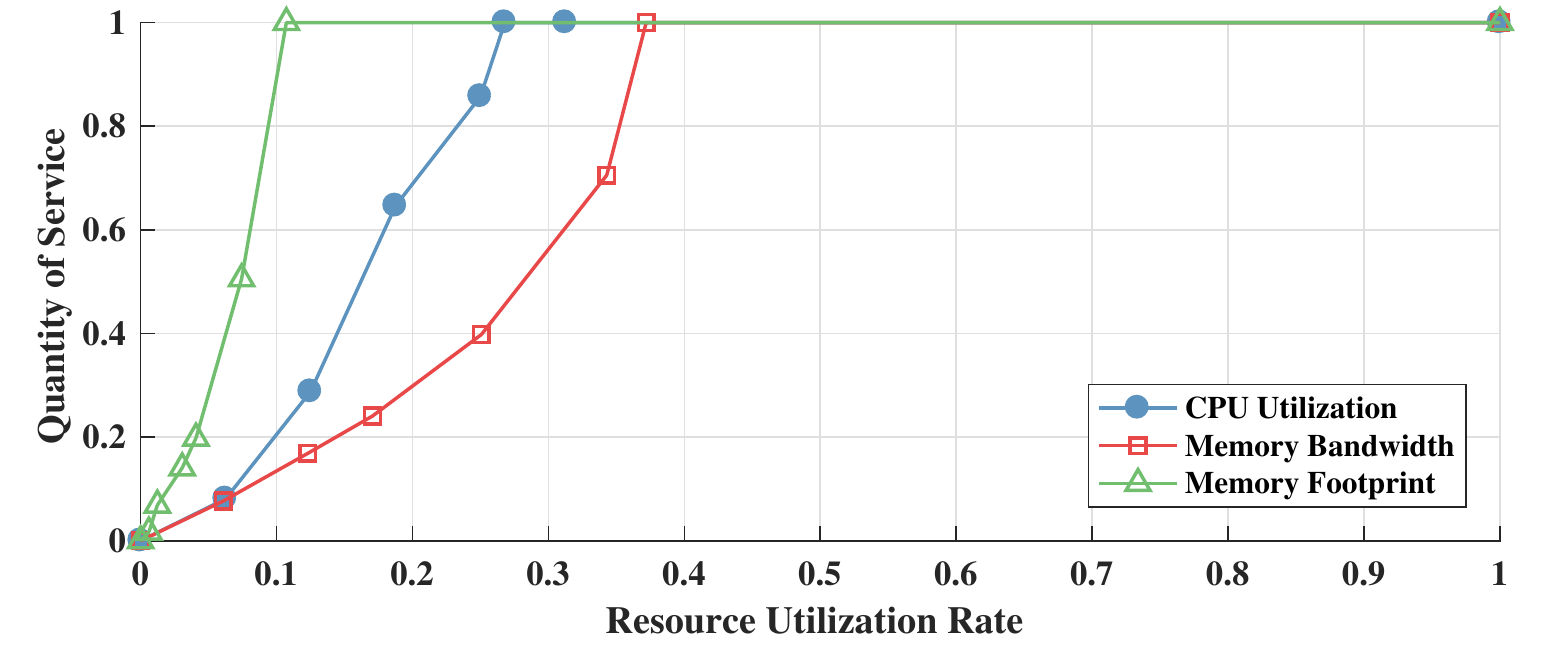}}
    \subfigure[\label{fig:curve_ssd}SSD]{\includegraphics[height=1.3in]{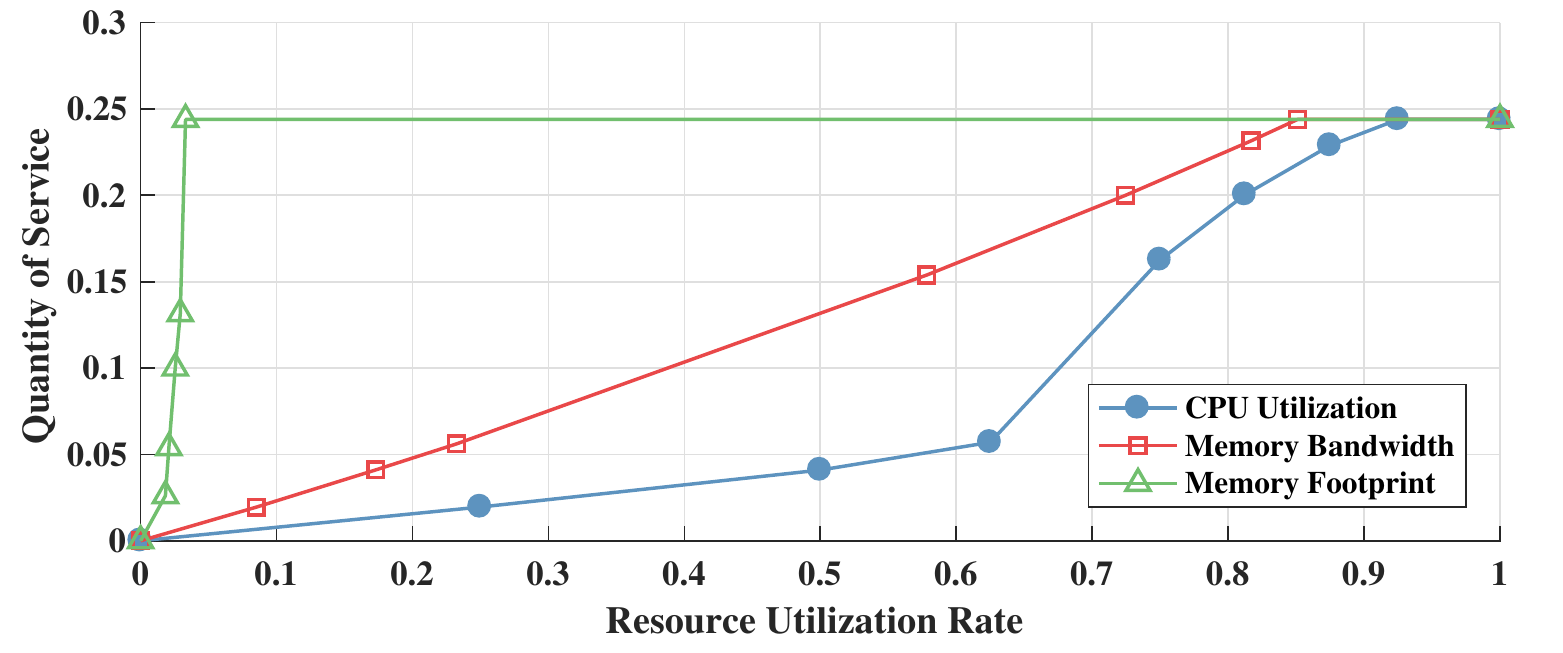}}
    \caption{QoS-RU Curves Case Study of ORB-SLAM2 and SSD.}
    \label{fig:curve}
\end{figure}

The application perspective metric give the user an overview of the platform performance when running different applications. Furthermore, CAVBench provides the system perspective metric (QoS-RU) to calculate a quantitative benchmarking result. We take ORB-SLAM2 and SSD as the case study. According to the above results, we set $10$ FPS as the best QoS (QoS=1). 
The QoS-RU curves of ORB-SLAM2 and SSD are shown in Figure \ref{fig:curve}. As for ORB-SLAM2, the area under CPU utilization (CPU), memory bandwidth (MEMBAND) and memory footprint (MEMFOOT) curve are $0.84$, $0.75$, and $0.93$, respectively. The MF is $0.80$. In SSD case, the area under CPU, MEMBAND, and MEMFOOT curves are $0.09$, $0.14$, and $0.23$, respectively, and the MF is $0.15$. The results correspond to the application perspective metric, but it is from the system view and quantitative.

No doubt, if an application cannot reach the acceptable QoS, the MF will be poor. Meanwhile, if the application needs more system resources to reach high QoS, the MF will decrease more. In the real production environments, the on-vehicle applications run concurrently in one system and compete for the system's resources. Hence, the high QoS application with less system resource consumption will be preferable to CAVs computing systems. That is why we consider system resource utilization when calculating the Match Factor.

\section{Conclusion}
\label{sec:6conclu}
CAVBench is the first benchmark suite for computing system and architectures designed for connected and autonomous vehicles targeting computational performance evaluation. We chose four typical and dominate application scenarios of CAVs, and summarized six applications in these scenarios as the evaluation applications. After that, we collected state-of-the-art implementation and standard input datasets for each application and determined the output metrics of the CAVBench. We got three basic features from CAVBench. First, the CAVs application types are diverse, and the real-time applications are dominated in CAVs scenarios. Second, the input data is mostly unstructured. Third, the end-to-end deep learning algorithm is more preferable for CAVs computing system. Then, we ran a series experiments to explore the characteristics of CAVBench, and concluded three observations as follows. First, the operation intensity of the applications in CAVBench is polarized. Second, the applications in CAVBench all consume high memory bandwidth. Third, CAVBench has a lower cache miss rate on average, but for specific applications, the optimization of the cache architecture and data/instruction locality is still important. According to these features and characteristics, we presented some insights and suggestions about the CAVs computing system design or CAVs application implementation. Finally, we used the CAVBench to evaluate a typical edge computing platform and presented the quantitative and qualitative analysis of the benchmarking results.

We hope this work will be helpful to researchers and developers who target the computing system or architecture design of connected and autonomous vehicles. According to the insights proposed in this paper, our future work will proceed from the following aspects. First, we will focus on providing the CUDA and OpenCL implementation of the CAVBench to support more heterogeneous platforms. Second, we will explore more methodologies to evaluate the computing system in the CAVs scenarios comprehensively, such as a benchmark dedicating system memory behaviors. Third, we will implement a full stack computing system for CAVs that will be competent for all CAVs applications.

\section*{Acknowledgment}
The authors are very grateful to the reviewers and our shepherd Fan Bai for their constructive comments and suggestions. We also would like to thank our colleague Yongtao Yao, who provided the implementation of some applications in CAVBench. This work is supported in part by National Science Foundation (NSF) grant CNS-1741635. 

\bibliographystyle{IEEEtran}
\bibliography{cav_system,cav_algorithm,benchmark,edge,other}

\end{document}